\documentstyle[epsfig]{mn2e}

\title[High speed photometry of faint CVs - VI.]
{High speed photometry of faint cataclysmic variables - VI.  Car2, 
V1040 Cen, H$\alpha$075648, IL Nor (Nova Nor 1893), HS Pup (Nova Pup 1963), 
SDSS\,J2048-06, CSS\,081419-005022 and CSS\,112634-100210}
\author[Patrick A.~Woudt and Brian Warner]
       {Patrick A.~Woudt$^{1}$\thanks{email: Patrick.Woudt@uct.ac.za},
Brian Warner$^{1,2}$\thanks{email: Brian.Warner@uct.ac.za}\\
        $^1$ Department of Astronomy, University of Cape Town, Private Bag X3,
        Rondebosch 7701, South Africa\\
        $^2$ School of Physics and Astronomy, Southampton University, Highfield, 
        Southampton SO17 1BJ, UK}
\date{Accepted 2009 November 27. Received 2009 November 24; in original form 2009 April 22}

\begin{document}

\maketitle

\begin{abstract}
We have observed 8 faint cataclysmic variable stars photometrically. 
The nova-like Car2 was extensively sampled but showed little variability.
V1040 Cen was observed near the end of a dwarf nova outburst and possessed dwarf nova and 
quasi-periodic oscillations. H$\alpha$075648 has strong large amplitude flickering and 
a possible orbital modulation at 3.49 h. The correct identification for the 
nova remnant IL Nor (Nova Nor 1893) has been established. HS Pup (Nova Pup 1963) 
has a possible orbital period of 3.244 h. 
SDSS J2048-06 is a low mass transfer dwarf nova that in 
quiescence shows slow variations at 7.67 h 
(though poorly sampled with our observations) and an orbital modulation at 87.26 min. The dwarf nova 
CSS\,081419-005022 has an orbital period of 1.796 h and the eclipsing dwarf nova 
CSS\,112634-100210 has an orbital period of 1.8581 h.

\end{abstract}


\section{Introduction}

In five previous papers (Woudt \& Warner 2001, 2002, 2003; Woudt, 
Warner \& Pretorius 2004; Woudt, Warner \& Spark 2005a) we have presented 
results of high speed photometry of faint cataclysmic variables (CVs: see 
Warner 1995 for a review) that had only previously been poorly studied. Here 
we give light curves and analyses of a further 8 stars. We again emphasize that 
this work is in the nature of a survey -- many of the stars will require 
further study.

As with the earlier work we used the University of Cape Town  
CCD photometer, as described by O'Donoghue (1995), in frame transfer mode and 
with white light, on the 1.9-m (74-in) and 1.0-m (40-in) reflectors at the 
Sutherland site of the South African Astronomical Observatory. Our 
magnitude scale was derived using hot white dwarf standards, but because of 
the non-standard spectral distributions of CVs and the use of white light 
our magnitudes approximate a V scale only to $\sim$ 0.1 mag.

In Section 2 we give the results of our observations. Section 3 gives brief 
conclusions from our work.

\section{Observations}

\begin{table*}
 \centering
  \caption{Observing log.}
  \begin{tabular}{@{}llrrrrrcc@{}}
 Object       & Type         & Run No.  & Date of obs.          & HJD of first obs. & Length    & $t_{in}$ & Tel. &  V \\
              &              &          & (start of night)      &  (+2450000.0)     & (h)       &     (s)   &      & (mag) \\[10pt]
{\bf Car2}    &  NL     & S7476 & 2004 Nov 05 & 3315.57379 & 1.01 &   6 & 74-in & 15.4 \\
              &         & S7581 & 2005 Feb 18 & 3420.39242 & 0.25 &   6 & 74-in & 15.4 \\
              &         & S7587 & 2005 Feb 21 & 3423.28830 & 7.92 &   6 & 74-in & 15.4 \\
              &         & S7779 & 2008 Jan 20 & 4486.28983 & 0.46 &   6 & 74-in & 15.5 \\
              &         & S7780 & 2008 Jan 21 & 4487.27784 & 7.64 &   6 & 74-in & 15.5 \\
              &         & S7781 & 2008 Jan 22 & 4488.29017 & 7.41 &   6 & 74-in & 15.5  \\[5pt]
{\bf V1040 Cen} &  DN     & S7800 & 2008 Mar 12 & 4538.32689 & 1.66 &   6 & 74-in & 14.2 \\
              &         & S7802 & 2008 Mar 13 & 4539.29559 & 2.62 &   6 & 74-in & 14.0 \\
              &         & S7805 & 2008 Mar 14 & 4540.29887 & 2.31 &   5 & 74-in & 14.6 \\[5pt]
{\bf H$\alpha$075648}  &  CV     & S7784 & 2008 Feb 14 & 4511.29649 & 6.07 &  15 & 74-in & 16.5 \\
              &         & S7787 & 2008 Feb 15 & 4512.35407 & 3.75 &  15 & 74-in & 16.3 \\
              &         & S7788 & 2008 Feb 16 & 4513.27629 & 4.80 &  20 & 74-in & 16.4 \\
              &         & S7790 & 2008 Feb 17 & 4514.32668 & 5.20 &  15 & 74-in & 16.3 \\
              &         & S7792 & 2008 Feb 18 & 4515.26944 & 1.56 &  15 & 74-in & 16.3 \\[5pt]
{\bf IL Nor}  &  NR     & S6880 & 2003 Mar 29 & 2728.41370 & 1.08 &  30 & 74-in & 18.5: \\
              &         & S7263 & 2004 Feb 23 & 3059.54351 & 2.08 &  30 & 74-in & 19.0 \\[5pt]
{\bf HS Pup}  &  NR     & S6169 & 2000 Dec 01 & 1911.44058 & 0.45 &  15 & 74-in & 17.8 \\
              &         & S6190 & 2001 Feb 22 & 1963.26422 & 4.77 &  60 & 40-in & 17.8 \\
              &         & S7774 & 2008 Jan 16 & 4482.30724 & 6.89 &  45 & 74-in & 18.0 \\
              &         & S7775 & 2008 Jan 17 & 4483.30199 & 7.04 &  45 & 74-in & 18.0 \\
              &         & S7777 & 2008 Jan 18 & 4484.43505 & 3.91 &  45 & 74-in & 18.0 \\
              &         & S7778 & 2008 Jan 19 & 4485.30488 & 7.14 &  45 & 74-in & 18.0 \\[5pt]
{\bf SDSS\,2048}&  CV     & S7039 & 2003 Aug 20 & 2872.36263 & 3.35 &  90 & 40-in & 19.8 \\
              &         & S7051 & 2003 Aug 23 & 2875.37903 & 1.48 &  60 & 40-in & 20.0 \\
              &         & S7373 & 2004 Aug 10 & 3228.32200 & 5.82 &  60 & 74-in & 19.9 \\
              &         & S7376 & 2004 Aug 11 & 3229.32626 & 5.67 &  60 & 74-in & 19.8 \\
              &         & S7380 & 2004 Aug 12 & 3230.26432 &7.21  &60, 90& 74-in & 19.8 \\
              &         & S7382 & 2004 Aug 13 & 3231.44414 & 0.54 &  90 & 74-in & 19.8 \\
              &         & S7383 & 2004 Aug 16 & 3234.28729 & 1.60 &  90 & 74-in & 20.1 \\
              &         & S7385 & 2004 Aug 17 & 3235.43276 & 3.03 & 120 & 40-in & 19.9 \\
              &         & S7387 & 2004 Aug 18 & 3236.38654 & 2.74 & 120 & 40-in & 19.8 \\
              &         & S7390 & 2004 Aug 19 & 3237.33275 & 3.23 & 120 & 40-in & 19.7 \\
              &         & S7392 & 2004 Aug 21 & 3239.40634 & 1.93 & 120 & 40-in & 19.7 \\
              &         & S7395 & 2004 Aug 22 & 3240.35944 & 2.80 & 120 & 40-in & 19.8 \\
              &         & S7398 & 2004 Aug 23 & 3241.43877 & 2.37 & 120 & 40-in & 19.8 \\
              &         & S7439 & 2004 Sep 14 & 3263.30688 & 3.47 & 120 & 40-in & 19.8 \\
              &         & S7441 & 2004 Sep 15 & 3264.24061 & 5.53 & 120 & 40-in & 19.8 \\
              &         & S7444 & 2004 Sep 17 & 3266.28454 & 1.03 & 120 & 40-in & 20.0 \\
              &         & S7446 & 2004 Sep 19 & 3268.23968 & 4.57 & 120 & 40-in & 19.9 \\
              &         & S7448 & 2004 Sep 20 & 3269.24304 & 4.41 & 120 & 40-in & 19.8 \\[5pt]
{\bf CSS\,0814} &  DN     & S7847 & 2009 Mar 20 & 4911.30915 & 3.30 &  15 & 74-in & 18.6 \\
              &         & S7850 & 2009 Mar 21 & 4912.24181 & 4.89 &  30 & 74-in & 18.7 \\
              &         & S7852 & 2009 Mar 22 & 4913.24178 & 4.51 &  30 & 74-in & 18.5 \\
              &         & S7860 & 2009 Mar 24 & 4915.29105 & 2.41 &  30 & 74-in & 18.5 \\[5pt]
{\bf CSS\,1126} &  DN     & S7838 & 2009 Mar 02 & 4893.44596 & 4.87 &  60 & 74-in & 18.2$^*$ \\
              &         & S7841 & 2009 Mar 03 & 4894.36824 & 5.05 &  60 & 74-in & 18.3$^*$ \\
              &         & S7844 & 2009 Mar 19 & 4910.48091 & 3.08 &  60 & 74-in & 18.3$^*$ \\
              &         & S7857 & 2009 Mar 23 & 4914.37147 & 2.65 &  60 & 74-in & 18.3$^*$ \\[5pt]
\end{tabular}
{\footnotesize 
\newline 
Notes: NL = Novalike; NR = Nova Remnant; DN = Dwarf Nova; $t_{in}$ is the integration time; $^*$ mean magnitude out of eclipse; `:' denotes an uncertain value.\hfill}
\label{woudttab1}
\end{table*}

\subsection{Carinae 2 (Car2)}

Car2, alias BPM 18764, although long recognized as an object of some interest, 
has not been dignified with a variable star designation because hitherto there 
was no evidence that it varies in brightness. It is listed as a white dwarf 
suspect by Eggen (1969), who made only one photometric measurement and hence did 
not detect its variability. Wickramasinghe and Bessell (1977) recognized its 
true character when they noticed broad, shallow absorption lines accompanied by 
emission cores in H$\alpha$ and H$\beta$, found it to be close to an X-Ray source, 
and thought it might be an old nova. The spectral emission features ensured that it 
entered the CV catalogue\footnote{http://archive.stsci.edu/prepds/cvcat/index.html} 
(Downes et al.~2001), designated as Car2. Kawka et al.~(2007) 
find that in fact there is no X-Ray candidate in the ROSAT database at the 
position of Car2, but from infrared colours deduced that it is probably a white 
dwarf with a cool companion, the latter having a spectral type M3-4 V. Hoard et 
al.~(2007) similarly find evidence of a companion, probably of type M2-3 V.

\begin{figure}
\centerline{\hbox{\psfig{figure=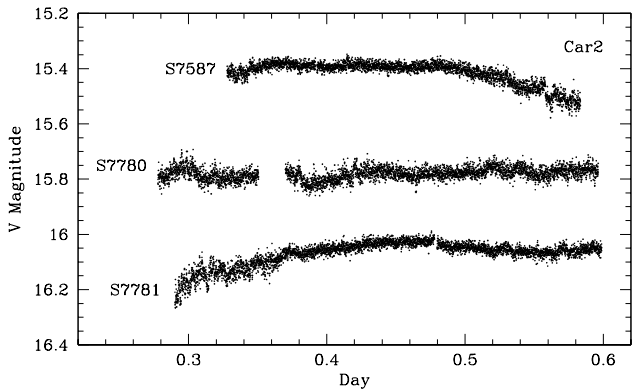,width=8.4cm}}}
  \caption{Individual light curves of Car2.  The light curve of run S7587 is displayed
at the correct brightness. Runs S7780 and S7781 have been 
displaced vertically by 0.3 and 0.6 mag, respectively, for display purposes.}
 \label{car2fig1}
\end{figure}

Our photometric runs are listed in Table~\ref{woudttab1} and the best three long 
light curves are shown in Fig.~\ref{car2fig1} (run S7587 is truncated because of poor
observing conditions). There are mean variations $\sim$ 0.1 mag from year to 
year, and some variations of similar amplitude during individual nights. A low 
amplitude cyclic variation may be suspected in the longer runs.
A Fourier transform (FT) of the combined January 2008 runs has a peak at 104.2 min which is probably
not significant.

\subsection{V1040 Centauri}

\begin{figure}
\centerline{\hbox{\psfig{figure=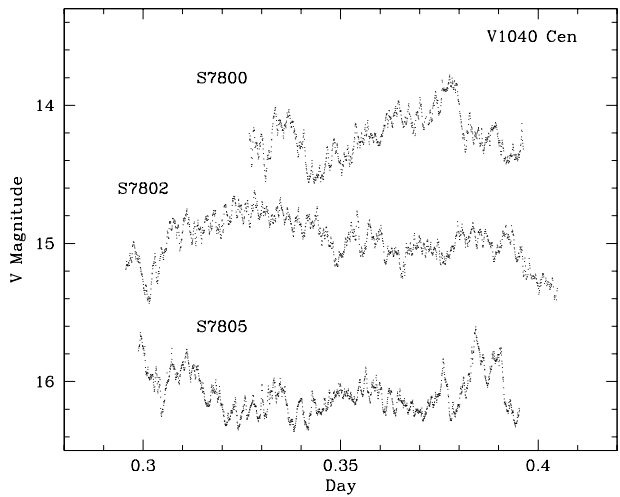,width=8.4cm}}}
  \caption{The 2008 light curves of V1040 Cen. Run S7800 is 
displayed at the correct brightness, runs S7802 and S7805 have
been displaced vertically by 1.0 and 1.5 mag, respectively,
for display purposes.}
 \label{v1040fig1}
\end{figure}

V1040 Cen was first found as an X-Ray source in the ROSAT Galactic Plane Survey 
(Motch et al 1998) and designated RX\,J1155.4-5641. It was classified as a CV, 
but no spectrum has been published. Photometry was carried out in a 
superoutburst in April 2002, and later in quiescence (Patterson et al.~2003), from 
which a superhump period of 1.492 h and an orbital period of 1.446 h were deduced, 
typical values for an SU UMa type dwarf nova. Recently Longa-Pe\~{n}a (2009) has found 
a spectroscopic period of 1.452 h. No other periods have been noticed. 

Our observing runs are listed in Table~\ref{woudttab1}. The AAVSO (American Association of
Variable Star Observers) light curve of V1040 Cen shows 
normal outbursts to V $\sim$ 12.3 with recurrence time $\sim$ 35 d and infrequent 
superoutbursts to V $\sim$ 11.6, reaching 10.8 on one occasion; Kato et al.~(2003) 
give the average time between superoutbursts as 211 d. At the time of our
observations V1040 Cen was returning to 
quiescence from a normal outburst that reached maximum about 4 days earlier 
than our first observation.

\begin{figure}
\centerline{\hbox{\psfig{figure=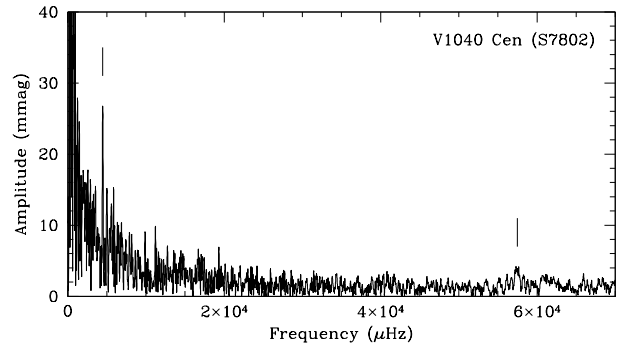,width=8.4cm}}}
  \caption{The Fourier transform of V1040 Cen (run S7802). The DNO and QPO signals at 17.4 s and 223 s, respectively,
have been marked by vertical bars.}
 \label{v1040fig2}
\end{figure}

Our light curves are shown in Fig.~\ref{v1040fig1}; the FTs contain only a few significant 
features, in particular that of run S7802 has a weak signal at 17.4 s and a 
stronger signal at 223 s -- see Fig.~\ref{v1040fig2}. The ratio of these is 13, which is similar 
to the ratio of QPO to DNO periods seen in dwarf novae during outburst 
(Warner, Woudt \& Pretorius 2003). In run S7800 there is a modulation at $\sim$ 225 s 
near the central part, which doubles to $\sim$ 455 s in the final third. Again, 
period doubling is commonly seen in DNO and QPO signals (Warner \& Woudt 2006).

\subsection{H$\alpha$\,075648}

\begin{figure}
\centerline{\hbox{\psfig{figure=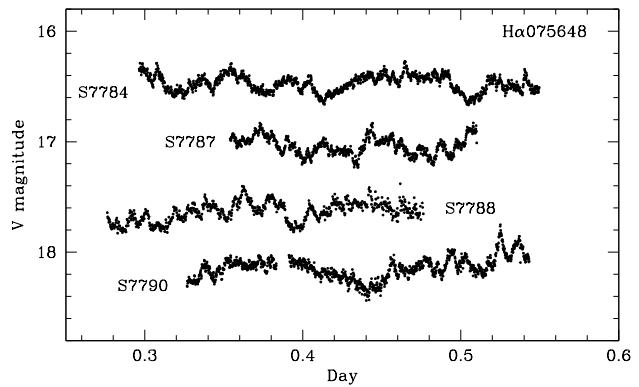,width=8.4cm}}}
  \caption{Individual light curves of H$\alpha$\,075648. The light curve of run S7784 is displayed
at the correct brightness. Runs S7787, S7788 and S7790 have been 
displaced vertically by 0.7, 1.2 and 1.8 mag, respectively, for display purposes.}
 \label{ha0756fig1}
\end{figure}

In a survey of CV candidates based on H$\alpha$ emission, Pretorius \& Knigge 
(2008) found 14 confirmed CVs and an additional two candidates. The B $\sim$ 18.3 
star H$\alpha$075648 showed very strong and rapidly variable H$\alpha$ emission 
but despite more than 7 h of spectroscopic coverage no plausible periodicity was 
detected. A short ($\sim$ 2 h) lightcurve obtained in poor conditions showed 
the expected flickering but no sign of orbital modulation. At the suggestion of 
Retha Pretorius we added H$\alpha$075648 to our observing list. Our photometric 
runs are listed in Table~\ref{woudttab1}, and the light curves are displayed in Fig.~\ref{ha0756fig1}.

  The star shows large scale flickering, which makes certain detection of any 
periodicity difficult, but an FT of the combined data set has a peak and window 
function at 3.49 h with an amplitude of 48 mmag. This is at best only 
a tentative suggestion of an orbital period.

\subsection{IL Normae (Nova Normae 1893)}

\begin{figure}
\centerline{\hbox{\psfig{figure=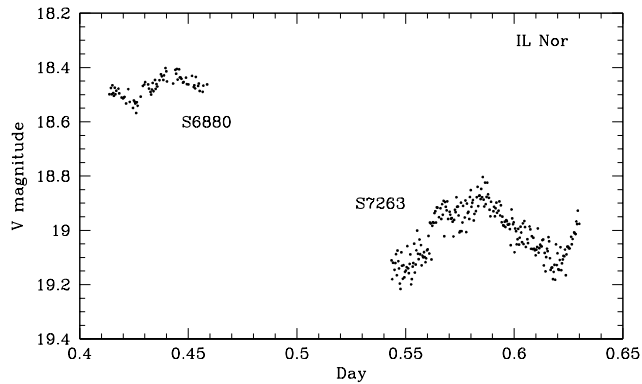,width=8.4cm}}}
  \caption{Light curves of IL Nor obtained on 2003 Mar 29 (S6880) and 2004 Feb 23 (S7263), respectively.}
 \label{ilnorfig1}
\end{figure}

\begin{figure}
\centerline{\hbox{\psfig{figure=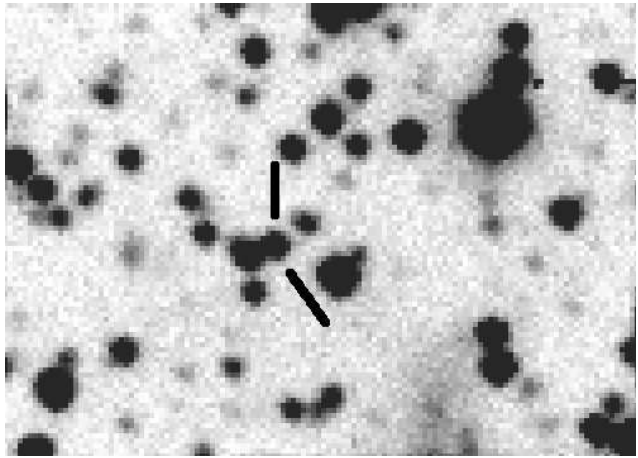,width=8.4cm}}}
  \caption{CCD image of IL Nor (indicated by the markers) obtained on 2004 February 23. 
The field of view is 50 by 34 arcsec, north is up and east is to the left.}
 \label{ilnorfig2}
\end{figure}

IL Nor was discovered in 1893 and reached a peak magnitude of 7.0 in July of 
that year (Duerbeck 1987). It was a moderately fast nova ($t_3$ = 108 d) which 
faded to less than 18th magnitude and has not been subsequently identified 
(Downes et al.~2001), being described by Duerbeck as coinciding with a blend 
of three stars, indicated on Duerbeck's chart and encircled in the Downes 
et al.~(2001) catalogue. To obtain good light curves for the components of the blend 
requires particularly good seeing, which we have achieved on only one occasion. 
Our observing runs are listed in Table~\ref{woudttab1} and two light curves are 
given in Fig.~\ref{ilnorfig1}, of which the first was able to show which component varies 
but only the second gave a satisfactory result. There is evidence for a 
modulation on a time scale $\sim$ 6000 s. A finding chart made from one 
of our CCD images is given in Fig.~\ref{ilnorfig2}. IL Nor was 18.5 mag in 2003 but $\sim$ 
0.5 mag fainter in 2004.

\subsection{HS Puppis (Nova Puppis 1963)}

HS Pup was the second nova to appear in the constellation of Puppis in 1963, 
reaching maximum at V = 8.0 on 23 December of that year (Strohmeier 1964). It was 
a moderately fast nova, decaying with $t_3$ = 65 d, and is classified as possibly 
of type B, meaning that the decline light curve (which was not well observed) may 
have had irregularities in it (Duerbeck 1981, 1987). The shell ejected during 
the nova explosion has been imaged (Gill \& O'Brien 1998) and the spectrum of 
the nova remnant shows H$\alpha$ in emission (Zwitter \& Munari 1995). HS Pup has no 
counterpart in the ROSAT All-sky Survey. 

The apparent magnitude of the remnant is variously given as $m_{pg}$ = 20.5 
(Duerbeck 1987), V = 18.06 (Szkody 1994), V = 19.1 (Zwitter \& Munari 1995) and 
J = 16.32 (Hoard et al.~2002). There is no published time-resolved photometry of 
this nova remnant.

\begin{figure}
\centerline{\hbox{\psfig{figure=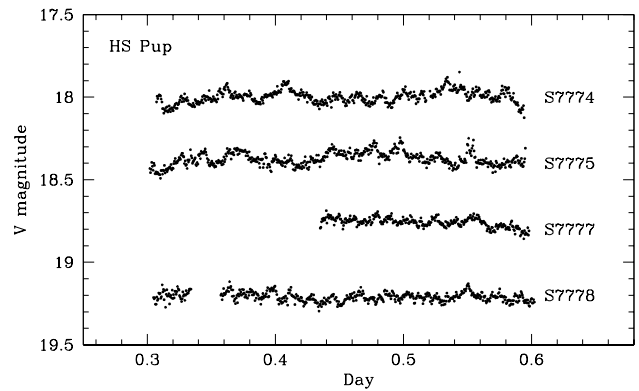,width=8.4cm}}}
  \caption{The 2008 light curves of HS Pup. The light curve of run S7774 is displayed
at the correct brightness. Runs S7775, S7777 and S7778 have been displaced vertically 
by 0.4, 0.8 and 1.2 mag, respectively, for display purposes.}
 \label{hspupfig1}
\end{figure}

   Our photometric observations of HS Pup are detailed in Table~\ref{woudttab1}. The 2008 light curves 
are shown in Fig.~\ref{hspupfig1}. There is a lot of flickering in the light curves of HS Pup, with
some excess power in the range of 600-700 $\mu$Hz ($\sim$ 1500 s). 
Despite extensive sampling of HS Pup with numerous 7-hour runs, there appears to be 
little coherency in the $\sim$ 1500 s modulation.
The FT of the combined 2008 observations, detrended by
subtracting the individual means, shows a marginal peak near 85.62 $\mu$Hz (3.244 h), with an amplitude
of 16.1 mmag. We tentatively assign this to an orbital frequency.

\subsection{SDSS\,J204817.85-061044.8}

SDSS 2048-06 was discovered during the Sloan Digital Sky Survey as a $g$ = 19.35 mag
CV with strong double hydrogen emission lines and indications of underlying 
broad absorption lines (Szkody et al.~2003). It is evidently a low $\dot{M}$ dwarf nova, 
but no outburst has yet been observed. No follow-up observations had been made 
before our own photometric work, which is listed in Table~\ref{woudttab1}, initial reports 
on which were published in Woudt et al.~(2005b) and Warner \& Woudt (2005).

\begin{figure}
\centerline{\hbox{\psfig{figure=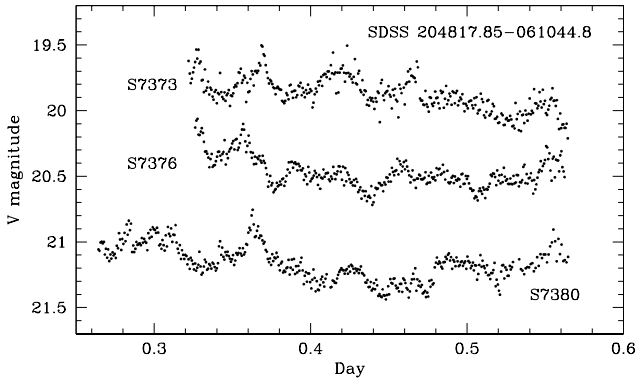,width=8.4cm}}}
  \caption{Individual light curves of SDSS\,2048-06. The light curve of run S7373 is displayed
at the correct brightness. Runs S7376 and S7380 have been 
displaced vertically by 0.7 and 1.4 mag, respectively, for display purposes.}
 \label{sdss2048fig1}
\end{figure}

\begin{figure}
\centerline{\hbox{\psfig{figure=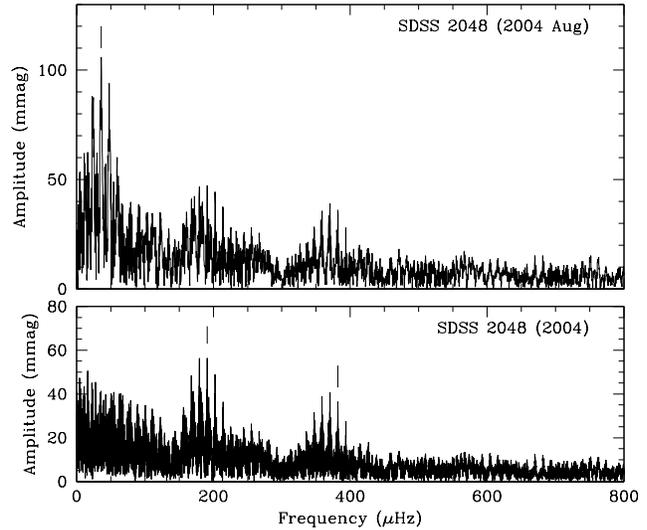,width=8.4cm}}}
  \caption{The Fourier transform of the 2004 August data (upper panel) and the 
combined 2004 August/September data (lower panel). The former is detrended by removing nightly
means. The 7.67-h periodicity is marked by the vertical bar. The lower panel shows 
data prewhitened with the 7.67-h periodicity;
the fundamental frequency of the orbital modulation and its first harmonic are marked by the 
vertical bars. }
 \label{sdss2048fig3}
\end{figure}

   A selection of long, individual light curves of the 2004 August observations 
is shown in Fig.~\ref{sdss2048fig1}. 
The FT of the August 2004 light curves is shown in 
upper panel of Fig.~\ref{sdss2048fig3} and shows a strong peak at 7.67 h with an amplitude of 112 mmag.
This periodicity is poorly sampled with the present data, given our data length
limitations (our three longest runs are 7.2, 5.8 and 5.7 hours, respectively) and should therefore
be regarded with caution. Nonetheless, it does not appear to be the result
of differential atmospheric extinction corrections, as the colours of SDSS2048-06 and the chosen
reference star are fairly similar and given identical airmass ranges on the three 
consecutive nights shown in Fig.~\ref{sdss2048fig1}.

After prewhitening the 2004 data with this periodicity, the FT, seen in 
the lower panel of Fig.~\ref{sdss2048fig3}, contains modulations consisting of a fundamental and its first 
harmonic modulation, but with an alias ambiguity. The two possibilities are 
$179.417 \pm 0.007$ $\mu$Hz (92.89 min) and 191.001 $\mu$Hz (87.26 min), both at 
amplitude 56 mmag. Comparison of separate FTs of our August and September 2004 
light curves gives a slight preference for the shorter of the two periods.
The orbital ephemeris for maximum light is

\begin{equation}
     {\rm HJD_{max}} = 245\,3228.3656 + 0.060597\,(2)\, {\rm E.}
\label{ephsdss2048}
\end{equation}

\begin{figure}
\centerline{\hbox{\psfig{figure=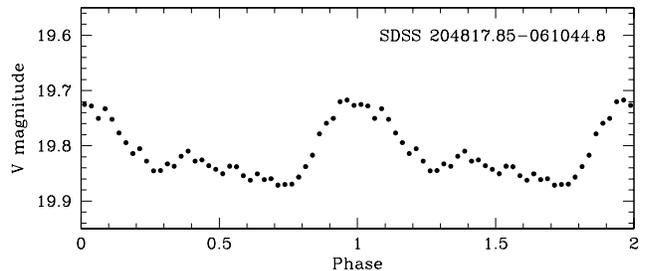,width=8.4cm}}}
  \caption{The average light curve of SDSS\,2048, folded on the ephemeris given in 
Eq.~\ref{ephsdss2048}.}
 \label{sdss2048fig4}
\end{figure}

The 87.26-min periodicity is almost certainly the orbital period -- its high stability 
over our 41 d baseline supports this. The mean light curve at the orbital period 
is shown in Fig.~\ref{sdss2048fig4} and may contain a grazing eclipse of the disc, superposed 
on an orbital hump.

   We first observed this star in the hope that it would have a non-radially 
pulsating white dwarf primary (e.g. Warner \& Woudt 2005), but we have found no 
evidence for such pulsations at a level of 1 mmag. Nor are there any signs of 
DNOs or QPOs.

\subsection{CSS\,081419-005022}

\begin{figure}
\centerline{\hbox{\psfig{figure=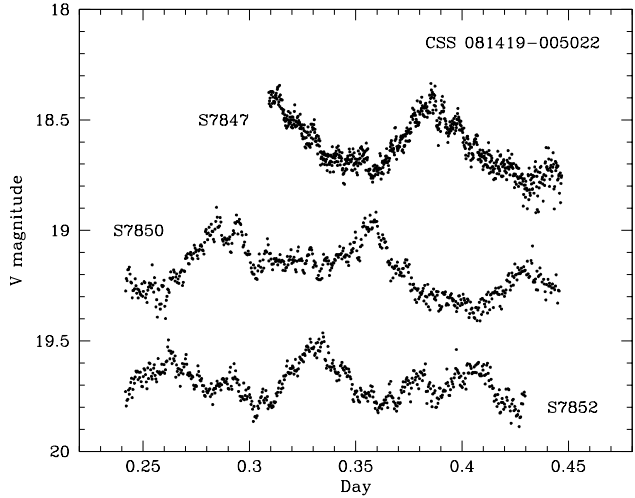,width=8.4cm}}}
  \caption{Individual light curves of CSS\,0814. The light curve of run S7847 is displayed
at the correct brightness. Runs S7850 and S7852 have been 
displaced vertically by 0.5 and 1.2 mag, respectively, for display purposes.}
 \label{css0814fig1}
\end{figure}

\begin{figure}
\centerline{\hbox{\psfig{figure=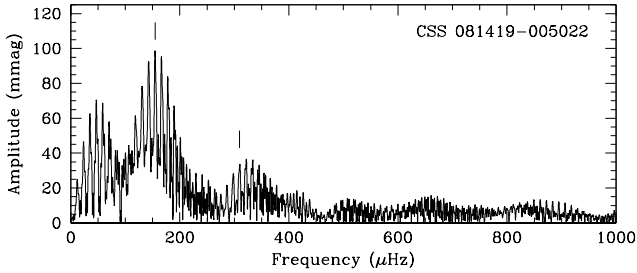,width=8.4cm}}}
  \caption{The Fourier transform of the combined runs S7847, S7850, S7852 and S7860 of CSS\,0814. 
The fundamental and first harmonic of the period quoted in the ephemeris (Eq.~\ref{ephcss0814})
are marked by the vertical bars.}
 \label{css0814fig2}
\end{figure}

\begin{figure}
\centerline{\hbox{\psfig{figure=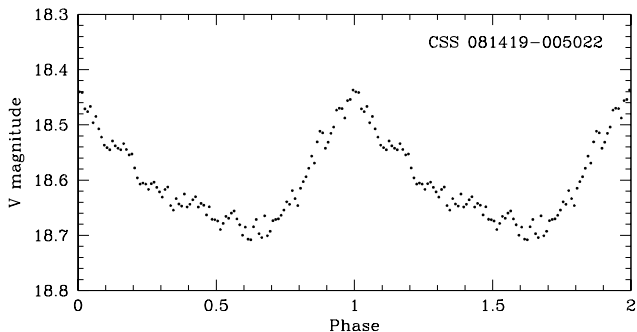,width=8.4cm}}}
  \caption{The average light curve of CSS\,0814, folded on the ephemeris given in 
Eq.~\ref{ephcss0814}.}
 \label{css0814fig3}
\end{figure}

This is the first of two CVs that we have observed, sourced from the Catalina 
Sky Survey (CSS)\footnote{See also http://nesssi.cacr.caltech.edu/catalina/AllCV.html} 
for near-earth objects (Drake et al.~2009), where it 
was listed as 19.0 at quiescence but with outbursts up to mag 14.8. It was 
listed in the SDSS survey but not recognized as a CV. Our observations are listed 
in Table~\ref{woudttab1} and the light curves of the three longest runs
are shown in Fig.~\ref{css0814fig1}. The FT of all observations (detrended by subtraction
individual means) is given in Fig.~\ref{css0814fig2}, in which we see a strong 
orbital modulation. A simultaneous 
sinusoidal fit of fundamental and first harmonic gives an orbital period of 
1.796 h at an amplitude of 98 mmag. The mean light curve at this period is 
shown in Fig.~\ref{css0814fig3}. The ephemeris for maximum of the orbital hump is

\begin{equation}
     {\rm HJD_{max}} = 245\,4911.385 + 0.07485\,(2)\, {\rm E.}
\label{ephcss0814}
\end{equation}

\subsection{CSS\,112634-100210}

\begin{figure}
\centerline{\hbox{\psfig{figure=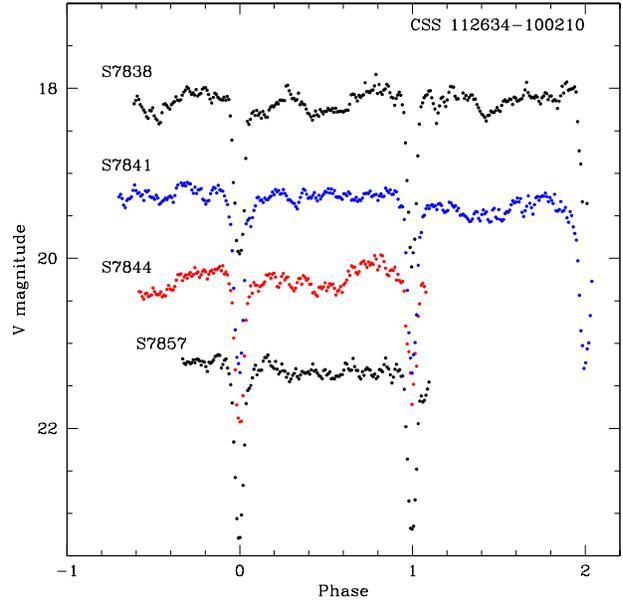,width=8.4cm}}}
  \caption{Individual light curves of CSS\,1126, folded on the emphemeris given in 
Eq.~\ref{ephcss1126}. The light curves of runs S7841, S7844 and S7857 have been 
displaced vertically for display purposes only.}
 \label{css1126fig1}
\end{figure}

This star is listed in the CSS at 18.6 mag  with a note that it is an eclipsing 
system. It was detected in the SDSS but no spectrum nor resulting identification 
as a CV was made. Our observations are listed in Table~\ref{woudttab1} and the light curves 
are shown in Fig.~\ref{css1126fig1}. There are deep eclipses, with period 1.8581 h, 
following the ephemeris for mid-eclipse:

\begin{equation}
     {\rm HJD_{min}} = 245\,4893.49325 + 0.077422\,(2)\, {\rm E.}
\label{ephcss1126}
\end{equation}

\begin{figure}
\centerline{\hbox{\psfig{figure=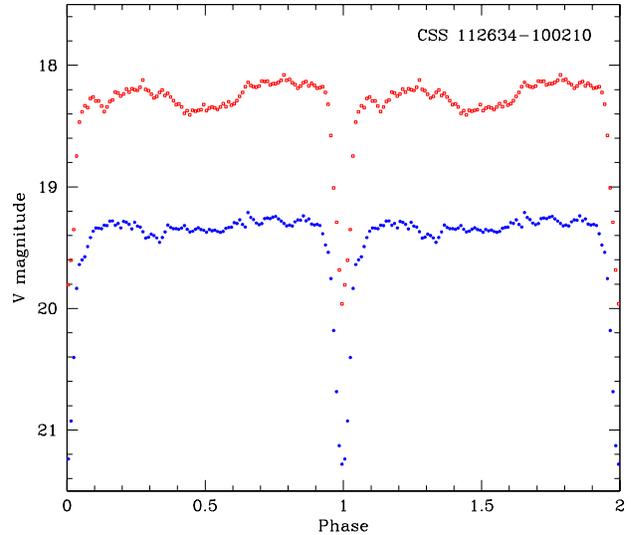,width=8.4cm}}}
  \caption{The average light curve of CSS\,1126. The top curve shows the average of runs S7838 and S7844,
the lower curve displays the average of runs S7841 and S7857, where the latter curve 
has been displaced vertically downwards by 1 mag for display purposes.}
 \label{css1126fig2}
\end{figure}

  We noticed that the orbital variations can be divided into two types of 
behaviour, one with almost no modulation out of eclipse and one with a double hump 
per orbit, in the style of the low $\dot{M}$ dwarf nova WZ Sge. We show these 
behaviours in Fig.~\ref{css1126fig2} as averages of runs S7841, S7857 and S7838, S7844 
respectively. The former have slightly deeper eclipses. There is little indication 
of any single hump per orbit contribution from a bright spot, and the profiles of 
the eclipses show only slight effects of a bright spot. 

\section{Conclusions}

The use of small or modest-apertured telescopes continues to be fruitful in 
the identification of CVs that will be worthy of follow-up observations with 
larger instruments. Our latest contribution includes our first photometric follow-up
of faint dwarf novae detected in the Catalina Sky Survey (CSS) demonstrating
the need for well-sampled photometric observations of transients detected in wide-area surveys. 
The two CSS dwarf novae presented here have orbital periods of 1.796 h (CSS 081419-005022) and 
1.8581 h (CSS 112634-100210), respectively, where the latter shows deep eclipses.

Further observations were presented of two nova remnants (HS Pup and IL Nor). 
The nova remnant IL Nor has been unambiguously identified, and extensive photometry
of HS Pup has resulted in a tentative identification of the orbital period in this
system (3.244 h).

The low mass transfer dwarf nova SDSS 2048-06 was extensively studied with the aim of
identifying non-radial pulsations of the accreting white dwarf primary. No pulsations were detected
to a level of 1 mmag, but a (poor sampled) long-period variability was seen ($\sim$ 7.67 h), and 
the orbital period has been securely identified (1.454 h).

\section*{Acknowledgments}

Both authors acknowledge support from the University of Cape Town and from the 
National Research Foundation of South Africa. We thank the anonymous referee
for useful comments which helped to improve this paper. PAW in addition thanks the 
School of Physics and Astronomy at Southampton University for financial 
assistance. We kindly acknowledge Denise Dale for obtaining some data of
SDSS\,2048-06.

\end{document}